\documentclass[showpacs,amsmath,amssymb,aps,showkeys,floatfix,prd,a4paper]{revtex4}
\usepackage[dvips]{graphicx}
\usepackage{dcolumn}
\usepackage{bm} 
\usepackage{epsfig}
\usepackage{amsfonts}
\usepackage{amssymb,amscd}
\usepackage{subfigure}
\usepackage{xcolor}
\usepackage{verbatim}

\def\pom{{I\!\!P}}

\newcommand{\Pelotas}{High and Medium Energy Group, Instituto de F\'isica e Matem\'atica,
             Universidade Federal de Pelotas\\
             Caixa Postal 354,  96010-900, Pelotas, RS, Brazil.}

\newcommand{\FURG}{Instituto de Matem\'atica, Estat\'istica e F\'isica, Universidade Federal de Rio Grande \\ { Av. Italia, km 8, Campus Carreiros}, 96203-900, Rio Grande, RS, Brazil.}

\begin{document}

\title{Double heavy quarkonium production in diffractive processes at the Run 2 LHC energy}
\author{C. Brenner Mariotto$^{1}$, V. P. Gon\c calves$^{2}$, R. Palota da Silva$^{2}$}
\affiliation{$^{1}$ \FURG\\ $^{2}$ \Pelotas}

\begin{abstract}
The double quarkonium production in single and double diffractive processes is investigated considering $pp$ collisions at the Run 2 LHC energy. Using the  
 nonrelativistic QCD (NRQCD) factorization formalism for the quarkonium production and the Resolved Pomeron model to describe the diffractive processes, we estimate the rapidity and transverse momentum dependencies of the cross sections for the $J/\Psi J/\Psi$ and $\Upsilon \Upsilon$ production. The contributions of the color-singlet and color-octet channels are estimated and predictions for the total cross sections in the kinematical regions of the LHC experiments are also presented. Our results demonstrate that the double quarkonium production in diffractive processes is not negligible and that its study can be useful to test the underlying assumptions present in the description of the single and double diffractive processes.
\end{abstract}


\maketitle

\section{Introduction}

The study of hadronic collisions at the LHC provides a unique environment for precise measurements of poorly understood phenomena. In particular, the study of quarkonium production at the LHC is expected to provide important insight for improving the theoretical description of its mechanism of production, which represents one of the long-standing problems of Quantum Chromodynamics (QCD) \cite{rev1}. During the last years, the study of the heavy quarkonium states in different reactions was a theme of intense interest, with distinct approaches  proposed to describe the factorization 
of the small- and large-distance contributions present in the quarkonium production \cite{review_nrqcd}. One of the possible frameworks is the Non-Relativistic QCD (NRQCD) formalism \cite{nrqcd}, which is a non-relativistic effective theory equivalent to QCD, which considers that the heavy quarkonium cross section receives contributions of the color singlet and color octet mechanisms.  In this formalism,  the cross section for the production of a heavy quarkonium state $H$ factorizes as  $\sigma (ab \rightarrow H+X)=\sum_n \sigma(ab \rightarrow Q\bar{Q}[n] + X) \langle {\cal{O}}^H[n]\rangle$, where the coefficients $\sigma(ab \rightarrow Q\bar{Q}[n] + X)$ are perturbatively calculated short distance cross sections for the production of the heavy quark pair $Q\bar{Q}$ in an intermediate Fock state $n$, which does not have to be color neutral.  The $\langle {\cal{O}}^H[n]\rangle$
are nonperturbative long distance matrix elements, which describe the transition of the intermediate $Q\bar{Q}$ in the physical state $H$ via soft gluon radiation. Such quantities are determined from experimental data, with their values being, unfortunately, strongly dependent of the fit procedures 
  \cite{bute,han_prl15,zhang_prl15}. Although the NRQCD formalism is able to describe a large set of experimental data, recent results indicate that the situation is far from being conclusive. Consequently, further tests for the color singlet and color octet mechanisms in NRQCD are still need{ed} to clarify several aspects present in the heavy quarkonium production.

In the last years the production of  quarkonium pairs became an alternative for the study of the underlying { production mechanism} (See e.g. Refs. 
  \cite{Kart,Hump,Vogt,Qiao,Li,Qiao_jpg,Ko,Bere,trunin,lans4,Li_relativistic,lans2,lans3,Sun,lans1}
). The recent theoretical and experimental advances  were strongly motivated by the large contribution of multiple parton interactions at the LHC energies \cite{MPI}.  With the increasing of the energy, the probability that one proton - proton { collision} leads to more than one scattering process also increases. As a consequence, a double quarkonium $H_1 H_2$ final state  can be generated in a single partonic scattering (SPS) process, with $\sigma(pp \rightarrow H_1 H_2 X) \propto \sigma(gg \rightarrow H_1 H_2)$,  as well as in a double parton scattering (DPS) process, where $\sigma(pp \rightarrow H_1 H_2 X) \propto \sigma(gg \rightarrow H_1) \times \sigma(gg \rightarrow H_2)$ 
\cite{Kom_double,Baranov1,Baranov2,david_nuc,rafal_double,kulesza},
 and in higher order multiple scattering processes \cite{denterria_triple,rafal_triple}. Recent results for the double $J/\Psi$ production indicate that the double scattering contribution is comparable to the single one. In contrast, the double $\Upsilon$ production is expected to be dominated by the SPS  mechanism (See e.g. \cite{Bere2}). Considering the values predicted for the cross sections and the large luminosity present in the Run 2 of the LHC, a detailed experimental analysis of the double quarkonium production is expected to be feasible, which will allow to improve our understanding of the heavy quarkonium production.

\begin{figure}[t]
\begin{tabular}{ccc}
\centerline{
{
\includegraphics[scale =.5]{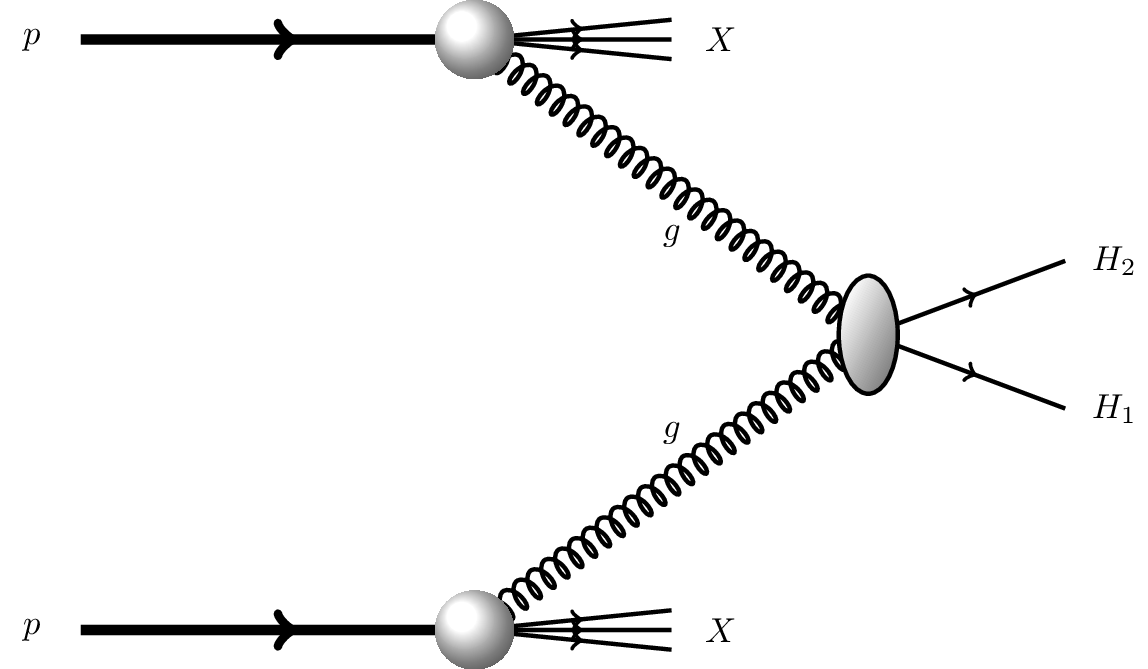}
}
{
\includegraphics[scale =.5]{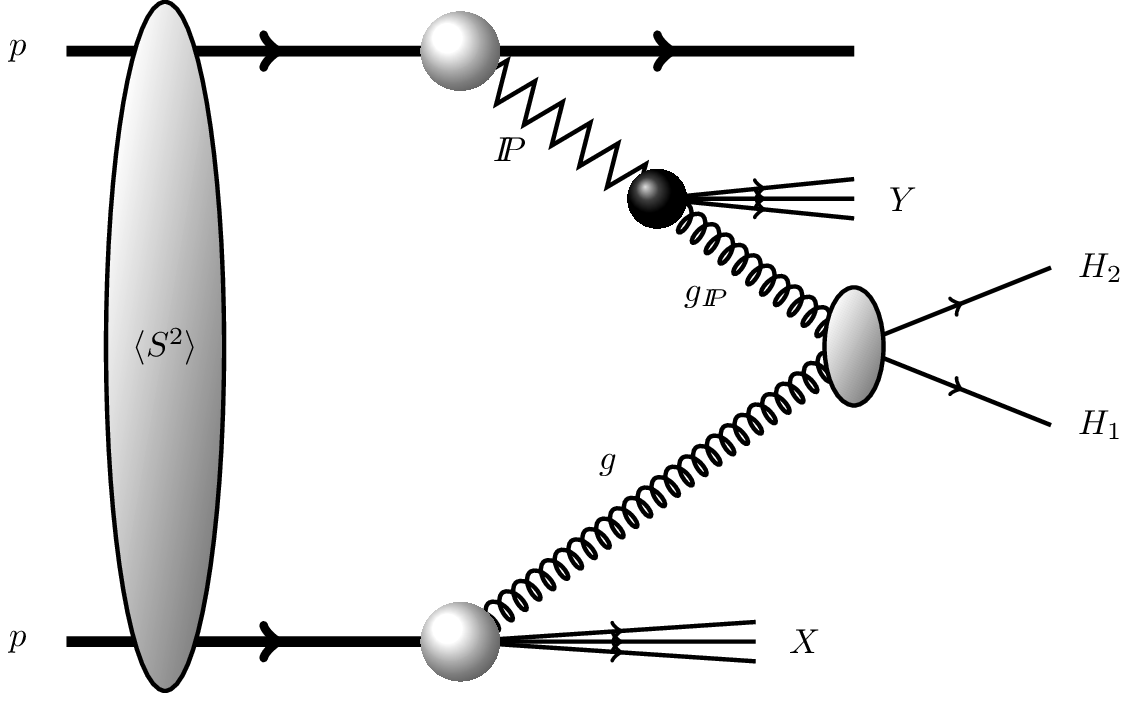}
}
{
\includegraphics[scale =.5]{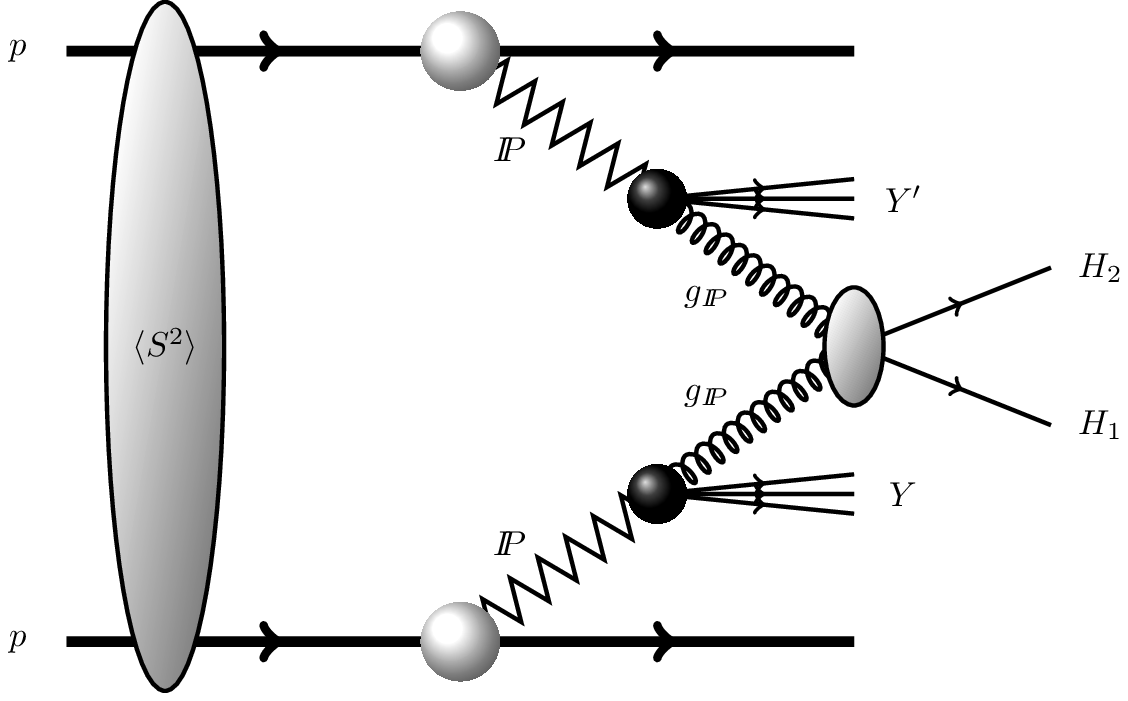}
}}
\end{tabular}
\caption{Typical diagrams for the double quarkonium production in inclusive (left panel), single diffractive (central panel) and double diffractive (right panel) processes. The blob denoted by  $\langle |S|^2\rangle$ in the single and double diffractive processes represents the gap survival factor associated to absorptive effects (See text for details). } 
\label{Diagramas}
\end{figure}

The recent studies of the double heavy quarkonium production in single parton scattering processes  are  in general dedicated to the calculation of the production of this final state in inclusive reactions, where both initial protons  dissociate in the interaction. A typical diagram is represented in Fig. \ref{Diagramas}(left panel). However, a double quarkonium  can also be produced in diffractive interactions, where one (or both) of the protons remain intact and   
empty regions  in pseudo-rapidity, called rapidity gaps, separate the intact very forward proton(s) from the $H_1 H_2$ state. Examples are the single and double diffractive processes, represented in the central and right panels of Fig. \ref{Diagramas}, respectively. 
Our motivation to study these processes is twofold. Firstly, the treatment of diffractive processes at the LHC have recently attracted much attention as a way of investigate the interplay of small- and large-distance dynamics within Quantum Chromodynamics \cite{review_forward}. Our goal is to complement previous studies \cite{khoze,Mariotto} considering other final states and present a comprehensive analysis of the  diffractive $J/\Psi J/\Psi$ and $\Upsilon \Upsilon$ production in $pp$ collisions at the Run 2 LHC energy ($\sqrt{s} = 13$ TeV). Second{ly}, the LHCb Collaboration have measured during the Run 1 the double charmonium production \cite{lhcb_double} in exclusive reactions, which are characterized by two rapidity gaps and nothing else is produced except the leading protons and the central charmonium states. New data, with larger statistics and smaller uncertainties, are expected to be released soon. However, in order to have access to these exclusive processes, it is fundamental to have control of the background associated to the production of this same final state in double diffractive processes.
{ The theoretical treatment of the exclusive double charmonium production was presented in Ref. \cite{khoze}. In contrast, in this paper we will focus on the treatment of  quarkonium pair production in double diffractive processes, which  also exhibit two rapidity gaps in the final state. However,} differently from the exclusive case,  they contain soft particles accompanying the production of a hard diffractive object (double quarkonium), with the rapidity gaps becoming,  in general, smaller than in the exclusive case.  The presence of the soft particles should increase the number of tracks in the detector. Therefore, in principle, the double diffractive and exclusive contributions could be separated using the exclusivity criteria. If this separation is feasible, is also allows to study in more detail the modelling of the diffractive processes and the  description of the quarkonium production. On the other hand, due to the large pile-up of events in each bunching crossing present in the Run 2, it is not clear if the separation of the  diffractive and exclusive events will be possible by measuring the rapidity gaps and counting the number of tracks in the final state. 
In principle, the only possibility to  detect  events characterized by two rapidity gaps, as those associated to the double diffractive and exclusive double quarkonium production, is by tagging the intact hadrons in the final state. It implies { that} the key element to measure diffractive events at the LHC will be tagging the forward scattered incoming hadrons \cite{detectors}. Moreover, if only one proton is tagged in the final state, the contribution of the double quarkonium production in single diffractive processes [Fig. \ref{Diagramas} (central panel)], should also be taken into account. { Predictions for the single diffractive production of a quarkonium pair are presented here for the first time.} Finally, it is important to emphasize that the DPS contribution for diffractive processes will be strongly suppressed, since the probability of two simultaneous interactions without the fragmentation of one (or both) of the incident proton is very small. The   typical topology of the  diffractive events (leading protons and rapidity gaps) is in general destroyed by multiple interactions. 
 All these aspects motivate the analysis that will be performed in what follows.

This paper is organized as follows. In the next Section we will review the formalism used to treat the double quarkonium production in single and double diffractive processes. The main aspects of the NRQCD formalism will be reviewed as well the basic assumptions of the  Resolved Pomeron Model will be presented. The treatment of the gap survival probability will  be also discussed. In Section \ref{res} we will present our predictions for the rapidity and transverse momentum distributions for the single and double diffractive $J/\Psi J/\Psi$ and $\Upsilon \Upsilon$ production in $pp$ collisions at the Run 2 LHC energy. Predictions for the total cross sections are also presented. A comparison with the predictions for the double quarkonium production in inclusive reactions is also performed. Finally, in Section \ref{conc} our main conclusions are summarized.

\section{Formalism}
\label{forma}

\begin{figure}[t]
\includegraphics[scale=1.2]{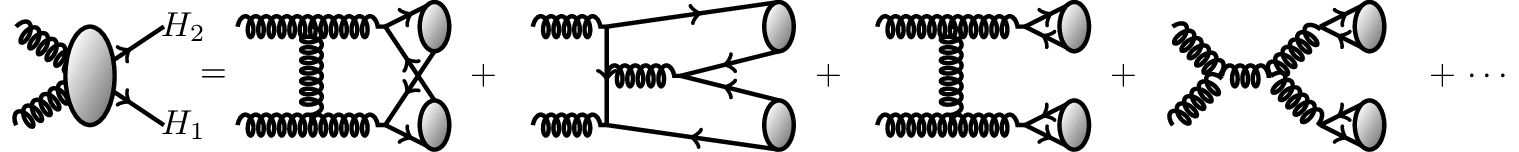}
\caption{Typical diagrams that contribute at leading order for the double quarkonium production by the gluon fusion subprocess.
}
\label{Fig:sub}
\end{figure}

In what follows we will present a brief review of the formalism needed to investigate the diffractive production of two spin-triplet S-wave quarkonia ($J/\Psi$ and $\Upsilon$) in $pp$ collisions at the Run 2 LHC energy. Such final states provide a clean experimental signature when studied in their decay into muons. In order to include the color-singlet and color-octet channels for each quarkonium final state, we will assume the NRQCD formalism \cite{nrqcd}, which takes into account of both contributions in a systematic way. In this formalism the cross section can be factorized as the product of short-distance and long-distance parts. The short-distance coefficients are process dependent, but can be calculated using perturbative QCD. In contrast, the long-distance matrix elements (LDMEs) are expected to be process independent, but are associated to nonperturbative physics. One important characteristic of the NRQCD formalism, is that a heavy quark pair, created in a color-octet state in the perturbative part { is also} allowed to contribute for the quarkonium production, with the color being neutralized in the long-distance regime. The { long}-distance contribution  can be expanded in terms of the { relative} velocity $v$ of the heavy quarks in the rest frame of the heavy quarkonium. Moreover, in the NRQCD formalism, the Fock states are expressed in terms of the dynamical gluons and the heavy quark pair, which implies that the wavefunction { of} a heavy quarkonium $H$ can be expressed as follows:
\begin{eqnarray}
|H\rangle = {\cal{O}}(1) |Q\bar{Q}_1 [^3S_1]\rangle + {\cal{O}}(v) |Q\bar{Q}_8 [^3P_J]g \rangle + {\cal{O}}(v^2) |Q\bar{Q}_{1,8} [^3S_1]gg \rangle + {\cal{O}}(v^2) |Q\bar{Q}_{8} [^3S_0]g \rangle + ... \,\,. 
\end{eqnarray}
As demonstrated in Refs. \cite{Ko,Qiao_jpg}, the dominant double quarkonium Fock states are the following combinations:
\begin{eqnarray}
|H_1\rangle |H_2\rangle & = & {\cal{O}}(1) |Q\bar{Q}_1 [^3S_1]\rangle |Q\bar{Q}_1 [^3S_1]\rangle + {\cal{O}}(v^4) |Q\bar{Q}_{8} [^3S_1]gg \rangle |Q\bar{Q}_{8} [^3S_1]gg \rangle  \,\,,
\end{eqnarray}
where the suppression present in the last term can be compensated by the enhancement of the large propagators present in the color octet channel for the production of a quarkonium pair { with large} transverse momentum. 
At the LHC energies, the double quarkonium production in inclusive and difffractive processes is dominated by the gluon fusion subprocess $gg \rightarrow H_1 H_2$. Some of the typical diagrams that contribute at leading order [${\cal{O}}(\alpha_s^4)$] for the cross section are presented in Fig. \ref{Fig:sub}. The  first two diagrams are examples 
of the $gg \rightarrow Q\bar{Q}_1(^3S_1) + Q\bar{Q}_1(^3S_1)$ subprocess, where the heavy quark pairs are produced in a color-{ singlet} state. In total, there are 31 diagrams that contribute for the color singlet channel. On the other hand, the color-octet channel, associated to the  $gg \rightarrow Q\bar{Q}_8(^3S_1) + Q\bar{Q}_8(^3S_1)$ subprocess, contributes with 72 diagrams, with two of them being represented by the last two diagrams in Fig. \ref{Fig:sub}.

In order to calculate the observables, the cross sections for the subprocesses, which are associated { with} the short-distance physics, should be multiplied by the large-distance matrix elements  $\langle O_1^{H}({}^3S_1) \rangle$ and 
$\langle O_8^{H}({}^3S_1) \rangle$. These matrix elements { account} for the transition probability of the heavy quark pair $Q \bar{Q}_n(^{2s+1}L_J)$ in a given color state $n$, spin $s$, angular momentum $L$ and total angular momentum $J$, to hadronize into quarkonium $H$. The color-singlet matrix elements $\langle O_1^{H}({}^3S_1) \rangle$ are usually determined from the leptonic decay of $H$. On the other hand, the color-octet one, $\langle O_8^{H}({}^3S_1) \rangle$, { are not known quite accurately} and has been obtained by fitting the transverse momentum spectrum for quarkonium production in a global analysis of different sets of experimental data. As a consequence, the predictions for the color-octet contribution are, in general, strongly dependent { on} the underlying assumptions for the associated matrix elements. Such uncertainty has direct impact in the predictions of the observables that are dominated by the color-octet channel. As we will show below, that is not the case { for} the double quarkonium production at small transverse momentum, which determines the magnitude of the total cross section. 

The leading order  color-singlet and color-octet contributions for the differential cross sections of the $gg \rightarrow H_1 H_2$ subprocess were calculated  by several authors in the nonrelativistic approximation  \cite{Kart,Hump,Vogt,Qiao,Li,Qiao_jpg,Ko,Bere} and more recently taking into { account the} relativistic effects \cite{trunin,Li_relativistic} and higher-order corrections \cite{lans4,lans3,Sun}. As the impact of the relativistic effects is still a theme of debate \cite{trunin,Li_relativistic} and the higher - order corrections are predicted to modify the transverse momentum distribution at large - $p_T$ \cite{lans4,lans3,Sun}, in our calculations for the diffractive production we will estimate the differential cross sections at leading order, disregarding these corrections. We will follow the notation from Refs. \cite{Qiao,Ko}, where the expressions for the differential cross sections are explicitly presented. In the case of the color-singlet contribution, the differential cross section, already multiplied by the corresponding matrix element, is given by \cite{Qiao}
\begin{eqnarray}
\frac{d\hat{\sigma}}{d\hat{t}}[gg\rightarrow  Q\bar{Q}_1({}^3S_1) Q\bar{Q}_1({}^3S_1)]\cdot \langle O_1^{H}({}^3S_1) \rangle^2 =
\frac{16\pi \alpha_s^4  |R_H(0)|^4}{81 M^2\hat{s}^8(M^2-\hat{t})^4(M^2-\hat{u})^4}\sum_{jkl} a_{jkl}M^j\hat{t}^k\hat{u}^l \,.
\label{csmdsdt}
\end{eqnarray}
where $|R_H(0)|^2$ is the square of the radial wave function of the quarkonium $H$ at the origin. The variables $\hat{s}$, $\hat{t}$ and $\hat{u}$ are the usual Mandelstam variables for the partonic subprocess, and in the nonrelativistic approximation we assume that $M=2m_Q$. For the charm and bottom quark masses we will { use} $m_c=1.5\,$ GeV and $m_b=4.7\,$ GeV.
The detailed expressions for the $a_{jkl}$ coefficients in Eq. (\ref{csmdsdt}) are given in Ref. \cite{Qiao}. Finally, the quarkonium radial function at the origin is related to the leptonic decay rate \cite{Eichten} as 
$\Gamma(H\to e^+e^-)=\frac{4N_c\alpha^2e_Q^2}{3}\frac{|R_H(0)|^2}{M_H^2}\left(1-\frac{16\alpha_s}{3\pi}\right)$\,. From recent PDG data \cite{pdg} 
for $\Gamma(J/\Psi\to e^+e^-)$ and $\Gamma(\Upsilon \to e^+e^-)$, we obtain $|R_{J/\Psi}(0)|^2=0.53\,$ GeV$^3$ and $|R_{\Upsilon}(0)|^2=4.6 \,$ GeV$^3$ for the $J/\Psi$ and the $\Upsilon$, respectively. For the color-octet contribution, the differential cross section, already multiplied by the corresponding matrix element, is given by \cite{Ko}
\begin{eqnarray}
\frac{d\hat{\sigma}}{d\hat{t}}[gg\rightarrow  Q\bar{Q}_8({}^3S_1)  Q\bar{Q}_8({}^3S_1)]
 \cdot \langle O_8^{H}({}^3S_1) \rangle^2 =
\frac{\pi^3 \alpha_s^4 \, \langle O_8^{H}({}^3S_1) \rangle^2}{972 M^6 \hat{s}^8(M^2-\hat{t})^4(M^2-\hat{u})^4}\sum_{j=0}^{14} a_j M^{2j} \,, 
\label{comdsdt}
\end{eqnarray}
where the $a_j$ coefficients can be found in Ref. \cite{Ko}. As in Ref. \cite{Ko} we will assume that  $\langle O_8^{J/\psi}({}^3S_1) \rangle=3.9\times 10^{-3} GeV^3$  \cite{Braaten2000} and $\langle O_8^{\Upsilon}({}^3S_1) \rangle=1.5\times 10^{-1} GeV^3$ \cite{kramer}. It is important to emphasize that more recent global analysis of the  world data for charmonium production \cite{bute} indicate that smaller values for $\langle O_8^{J/\psi}({}^3S_1) \rangle$ are prefered by the data. As consequence, our predictions for the associated color-octet contribution should be considered as an upper bound.

In order to estimate the double quarkonium production in single and diffractive  interactions we will assume the validity of the factorization theorem for these processes. Consequently, the hadronic cross sections will be given by the convolution of the  above differential cross sections with the inclusive and/or diffractive gluon distribution functions of the incident particles for the correspondent process. In the particular case of the single diffractive double quarkonium production, represented in Fig. \ref{Diagramas} (central panel),  the cross section can be expressed as follows   
\begin{eqnarray}
d\sigma (pp\rightarrow p + H_1H_2 + X) = \sum_n [g_{p} (x_1,\mu^2)\, g^D_{p} (x_2,\mu^2) + g^D_{p} (x_1,\mu^2) \, g_{p} (x_2,\mu^2)]  \cdot d\hat{\sigma}[gg\to Q\bar{Q}_n+Q\bar{Q}_n] \cdot \langle {\cal{O}}^{H_1}_{n} \rangle \langle {\cal{O}}^{H_2}_{n} \rangle \,\,,
\label{cssd}
\end{eqnarray}
with the final state being characterized by one intact hadron and one rapidity gap. In Eq. (\ref{cssd}) we take into account that any of the two incident protons can remain intact in the interaction. Moreover, $g_{p}$ and  $g^D$ are the inclusive and diffractive gluon distributions, probed at the scale $\mu^2$, which we assume to be equal to $\mu^2 = M^2 + p_T^2$. On the other hand, the cross section for the double diffractive process will be given by
\begin{eqnarray}
d\sigma (pp\rightarrow p + H_1H_2 + p) = \sum_n g^D_{p} (x_1,\mu^2)\, g^D_{p} (x_2,\mu^2)   \cdot d\hat{\sigma}[gg\to Q\bar{Q}_n+Q\bar{Q}_n] \cdot \langle {\cal{O}}^{H_1}_{n} \rangle \langle {\cal{O}}^{H_2}_{n} \rangle \,\,.
\label{csdd}
\end{eqnarray}
This process is represented in Fig. \ref{Diagramas} (right panel) and the final state will be characterized by two intact hadrons and two rapidity gaps. In our calculations we will { use}  
the CTEQ6L parametrization \cite{cteq} for the inclusive gluon distribution. For the diffractive gluon distribution, it will be modelled using the Resolved Pomeron Model \cite{IS}. The basic idea is that  the hard scattering resolves the quark and gluon content of the Pomeron \cite{IS} and  it can be obtained analysing the experimental data from diffractive deep inelastic scattering (DDIS) at HERA, providing us with the diffractive distributions of singlet quarks and gluons in the Pomeron \cite{H1diff}. Consequently,  the diffractive parton distributions are expressed in terms of parton distributions in the \,{Pomeron} and a Regge parametrization of the flux factor describing the \,{Pomeron} emission by the { proton}. The  parton distributions have evolution given by the DGLAP evolution equations and should be determined from events with a rapidity gap or a intact hadron. The diffractive gluon distribution, $g^D_{p} (x,Q^2)$, is defined as a convolution of the \,{Pomeron} flux emitted by the proton, $f^{p}_{\pom}(x_{\pom})$, and the gluon distribution in the \,{Pomeron}, $g_{\pom}(\beta, Q^2)$,  where $\beta$ is the momentum fraction carried by the partons inside the \,{Pomeron}.  The diffractive gluon distribution of the proton is then given by
\begin{eqnarray}
{ g^D_{p}(x,Q^2)}=\int dx_{\pom}~d\beta ~\delta (x-x_{\pom}\beta)~f^{p}_{\pom}(x_{\pom})~g_{\pom}(\beta, Q^2)={ \int_x^1 \frac{dx_{\pom}}{x_{\pom}} f^{p}_{\pom}(x_{\pom}) ~g_{\pom}\left(\frac{x}{x_{\pom}}, Q^2\right)} \,\,.
\label{difgluon:proton}
\end{eqnarray}
The \,{Pomeron} flux is given by
\begin{eqnarray}
f^{p}_{\pom}(x_{\pom})= \int_{t_{\rm min}}^{t_{\rm max}} dt \, f_{\pom/{p}}(x_{{\pom}}, t) = 
\int_{t_{\rm min}}^{t_{\rm max}} dt \, \frac{A_{\pom} \, e^{B_{\pom} t}}{x_{\pom}^{2\alpha_{\pom} (t)-1}}  \,\,,
\label{fluxpom:proton}
\end{eqnarray}
where $t_{\rm min}$, $t_{\rm max}$ are kinematic boundaries. The \,{Pomeron} flux factor is motivated by Regge theory, where the \,{Pomeron} trajectory is assumed to be linear, $\alpha_{\pom} (t)= \alpha_{\pom} (0) + \alpha_{\pom}^\prime t$, and the parameters $B_{\pom}$, $\alpha_{\pom}^\prime$ and their uncertainties are obtained from fits to H1 data  \cite{H1diff}. \,{The slope of the Pomeron flux is $B_{\pom}=5.5^{-2.0}_{+0.7}$ GeV$^{-2}$, the Regge trajectory of the Pomeron is
$\alpha_{\mathbb P}(t)=\alpha_{\mathbb P}(0)+\alpha_{\mathbb P}'~t$ with $\alpha_{\mathbb P}(0)=1.111 \pm 0.007$ and $\alpha_{\mathbb P}'=0.06^{+0.19}_{-0.06}$ GeV$^{-2}$. The $t$ integration boundaries are $t_{\rm max}=-m_{p}^2x_{\pom}^2/(1\!-\!x_{\pom})$ ($m_{p}$ denotes the proton mass) and $t_{\rm min}=-1$ GeV$^2$.} Finally, the normalization factor $A_{\mathbb P}=1.7101$ is chosen such that $x_{\pom}\times\int_{t_{\rm{min}}}^{t_{\rm{max}}}dt~f_{\pom/{p}}(x_{\pom},t)=1$ at $x_{\pom} = 0.003$.
In our analysis we use the fit B  obtained by the H1 Collaboration at DESY-HERA for the diffractive gluon distribution \cite{H1diff}. Finally, it is important to emphasize that the Resolved Pomeron model implies the presence of additional particles in the final state, associated to the remnants of the Pomeron, which dissociates in the interaction. As discussed in the Introduction, the presence of these particles can be used, in principle, to discriminate the diffractive from the exclusive double quarkonium production.

{
One important open question in the treatment of diffractive interactions in hadronic collisions is whether 
 the cross sections for the associated single and double diffractive processes are  somewhat modified by soft interactions which lead to an extra production of particles that destroy the rapidity gaps in the final state \cite{bjorken}. In the case of diffractive interactions in \,{${pp / p\bar{p}}$} collisions, the experimental results obtained at TEVATRON \cite{tevatron} and LHC \cite{atlas_dijet,cms_dijet} have demonstrated that one should take into { account these} additional absorption effects that imply the violation of the QCD hard scattering factorization theorem for diffraction \cite{collinsfac}. One { has} that the soft rescattering corrections associated to reinteractions (often
referred to as multiple scatterings) between spectator partons of the colliding hadrons produce additional final-state particles which fill the would-be
rapidity gap and suppress the diffractive events. Consequently, in order to estimate the diffractive cross sections in hadronic collisions we need to take into account { the} probability that such emission does not occur. The fact that the diffractive factorization breaking is intimately related to soft multiple scattering in hadron-hadron collisions has motivated the modelling of these effects using a general purpose Monte Carlo \cite{torbcris,qgsjetii,herwig}. A current shortcoming of these  promissing approaches is that, due to the complexity of the  diffractive interactions, their predictions are still strongly dependent on the treatment of the multiple interactions, the assumptions for the color flow along the rapidity gap as well as the modelling of possible proton excitations. Another possible approach to treat this problem is based on the assumption that the hard process occurs on a short enough timescale such that the physics that generate the additional particles can be factorized and accounted by an overall factor, denoted gap survival factor $\langle |S|^2\rangle$, multiplying the cross section calculated using the collinear factorization and the diffractive parton distributions extracted from HERA data. The modelling, magnitude and universality  of  this  factor {are still} a theme of intense debate \cite{review_martin,durham,telaviv}. In general the values of $\langle |S|^2\rangle$ depend on the energy, being typically of order 1 -- 5 \% for LHC energies. Such approach have been largely used in the literature to estimate the hard diffractive processes at the LHC (See e.g.  Refs. 
\cite{nosbottom,MMM1,antoni,antoni2,cristiano,kohara_marquet,nos_ze,nos_Dijet,nos_dimuons}) with reasonable success to describe the current data. However, as the effects that determine the gap survival { factor} have nonperturbative nature, they are difficult to treat and its magnitude is strongly model dependent (For recent reviews see Refs. \cite{durham,telaviv}).
In what follows we also follow this simplified approach, assuming $\langle |S|^2\rangle = 0.02$ for the double diffractive  production and $\langle |S|^2\rangle = 0.05$ for the single diffractive one. Such values were estimated  in Ref. \cite{kkmr}  using a two-channel eikonal formalism that take into account the contributions of high-mass diffractive dissociation, possible nucleon excitations in the diffractive interaction as well as the contribution of pion loops for the bare Pomeron pole. 
It is important to emphasize that this choice is somewhat arbitrary, and mainly motivated by the possibility to compare our predictions with those obtained in other analysis. Recent studies from the CMS Collaboration \cite{cms_dijet} indicate that this factor can be larger than this value by a factor $\approx 4$. Consequently, our results can be considered a lower bound for the diffractive contribution. However, it is important to emphasize that the uncertainty on  $\langle |S|^2\rangle$ only affect the normalization of the cross sections, with the shape of the distributions being a direct probe of the underlying assumption that the soft rescattering effects can be factorized of the hard process. In particular, if a different value for $\langle |S|^2\rangle$ is constrained by the experimental data for e.g.  diffractive heavy quark production, our predictions can be directly rescaled and  compared with the diffractive double quarkonium data. 

}

\begin{figure}
\begin{tabular}{c c}
\centerline{
{
\includegraphics[scale = .4]{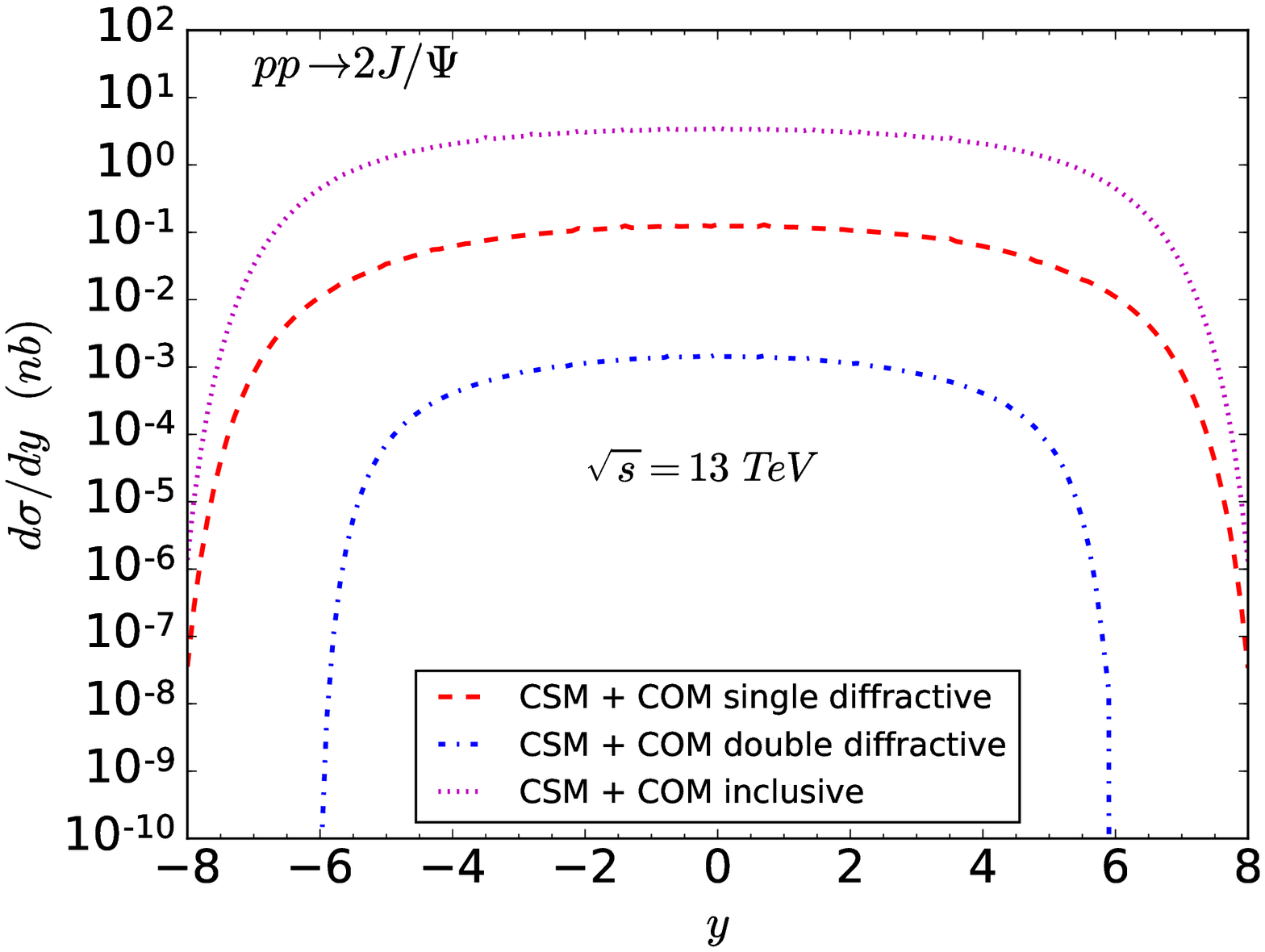}
}
{
\includegraphics[scale = .4]{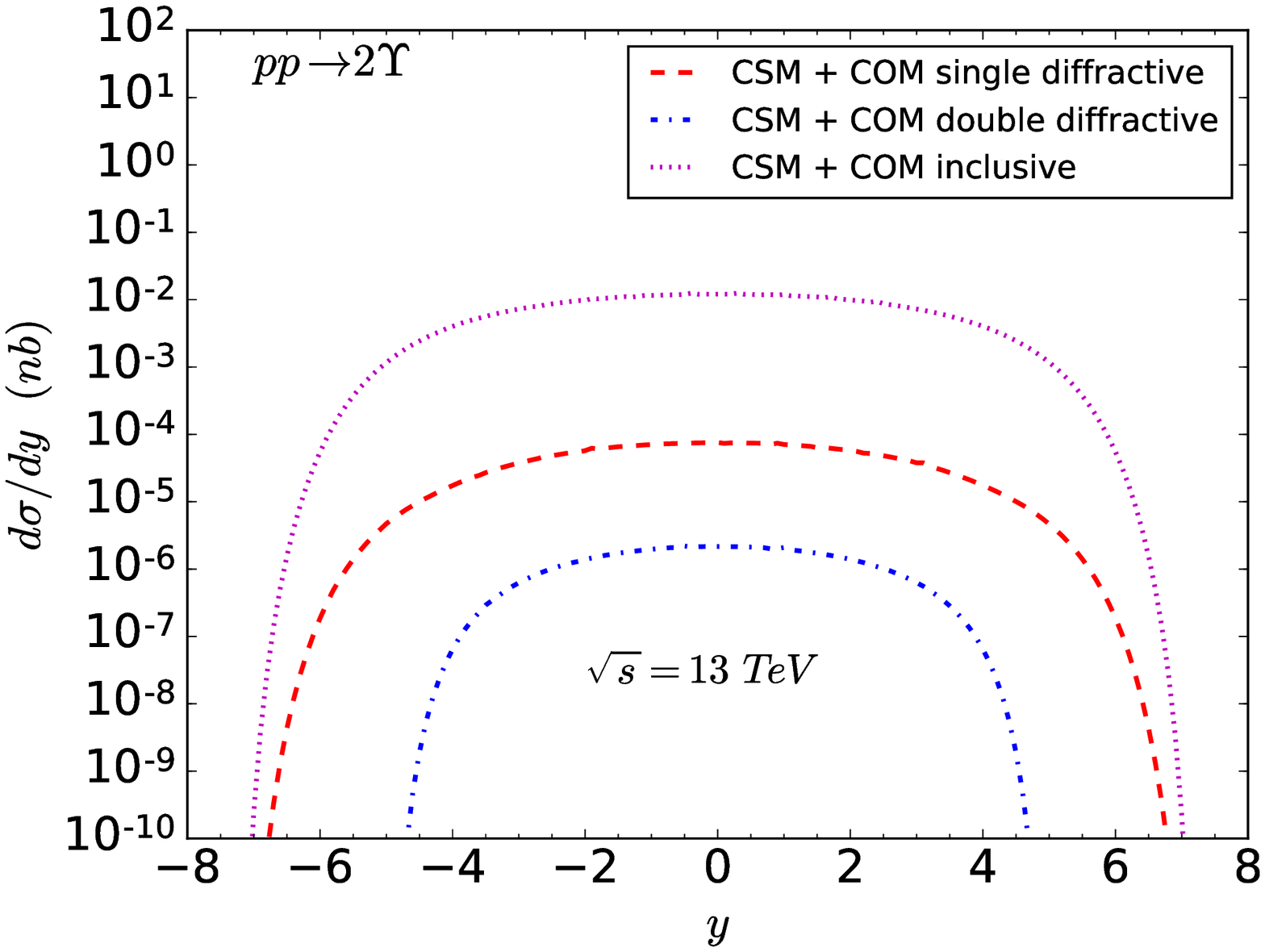}
}}
\end{tabular}
\caption{Rapidity distributions for double $J/\Psi$ production (left panel) and double $\Upsilon$  (right panel) production in single and double diffractive interactions. The predictions associated to the inclusive process are presented for comparison.} 
\label{Fig:Rapidez}
\end{figure}

\begin{table}[t]
\centering
\begin{tabular}{|l|c|c|c|}
\hline 
~ & {\bf Inclusive} & {\bf Single Diffrative (SD)} & {\bf Double Diffractive (DD)} \\ 
\hline
\hline 
${J/\Psi J/\Psi}$  ($|y| \le 8$) & 28.3 nb & $3.8\times 10^{-1}$ nb & $8.8\times 10^{-3}$ nb \\ 
\,\,\,\,\,\,\,\,\,\,\,\,\,\,\,\,\,\,\,\,\,\, ($2.0 \le y \le 4.5$) & 6.04 nb & $7.8\times 10^{-2}$ nb & $1.7\times 10^{-3}$ nb \\ 
\hline 
\hline 
${\Upsilon \Upsilon}$ ($|y| \le 8$) & 52.3 pb & $4.5\times 10^{-1}$ pb & $6.6\times 10^{-3}$ pb \\ 
\,\,\,\,\,\,\,\,\, ($2.0 \le y \le 4.5$) & 10.4 pb & $8.2\times 10^{-2}$ pb & $8.26\times 10^{-4}$ pb \\ 
\hline 
\end{tabular}
\caption{Predictions for the total cross sections for the double $J/\Psi$ and double $\Upsilon$ production in  single and double diffractive processes considering two rapidity ranges. The predictions for the inclusive production are presented for comparison.}
\label{tabela}
\end{table}

\begin{figure}[t]
\begin{tabular}{c c}
\centerline{
{
\includegraphics[scale = .4]{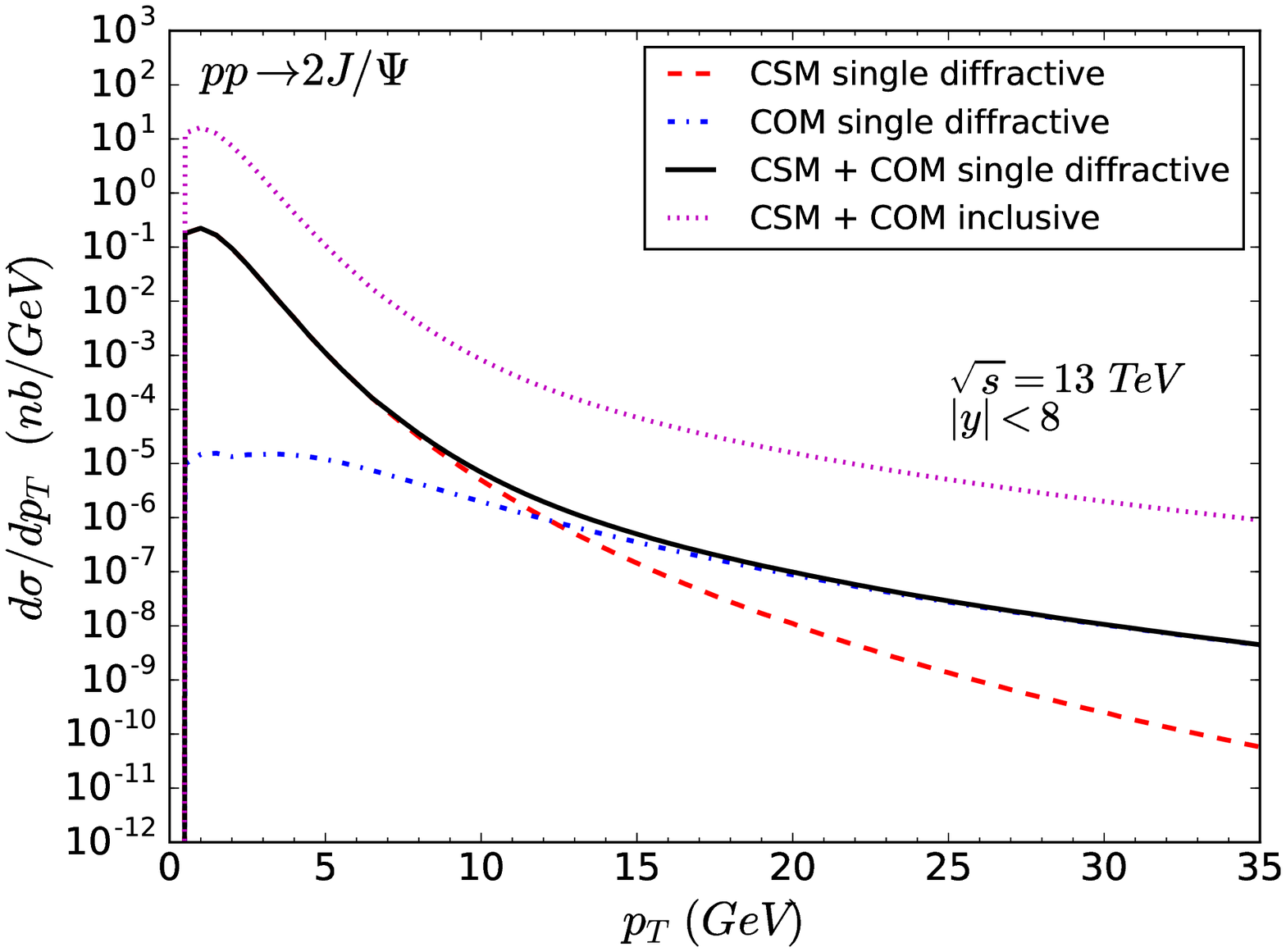}
}
{
\includegraphics[scale = .4]{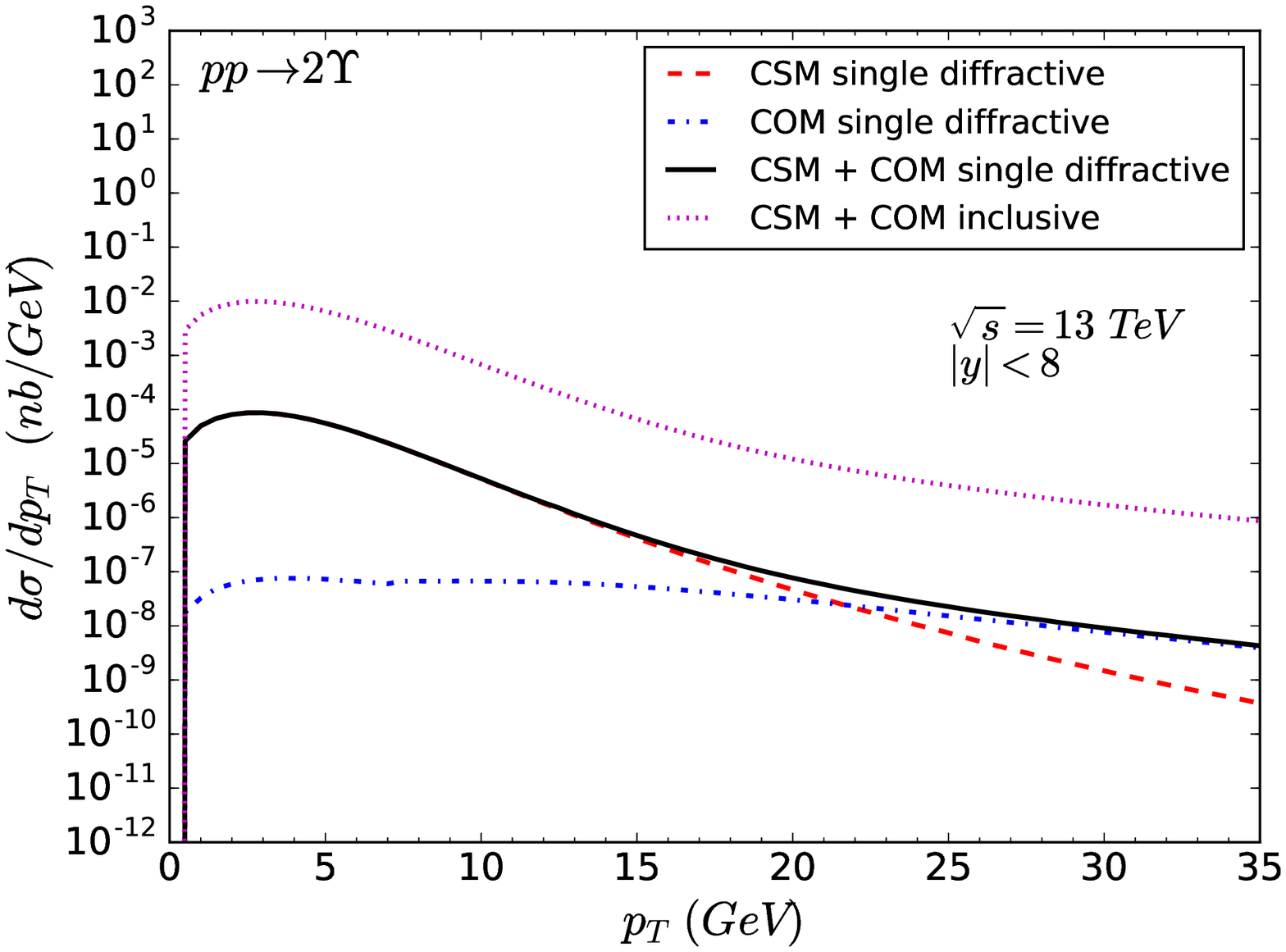}
}} \\
\centerline{
{
\includegraphics[scale = .4]{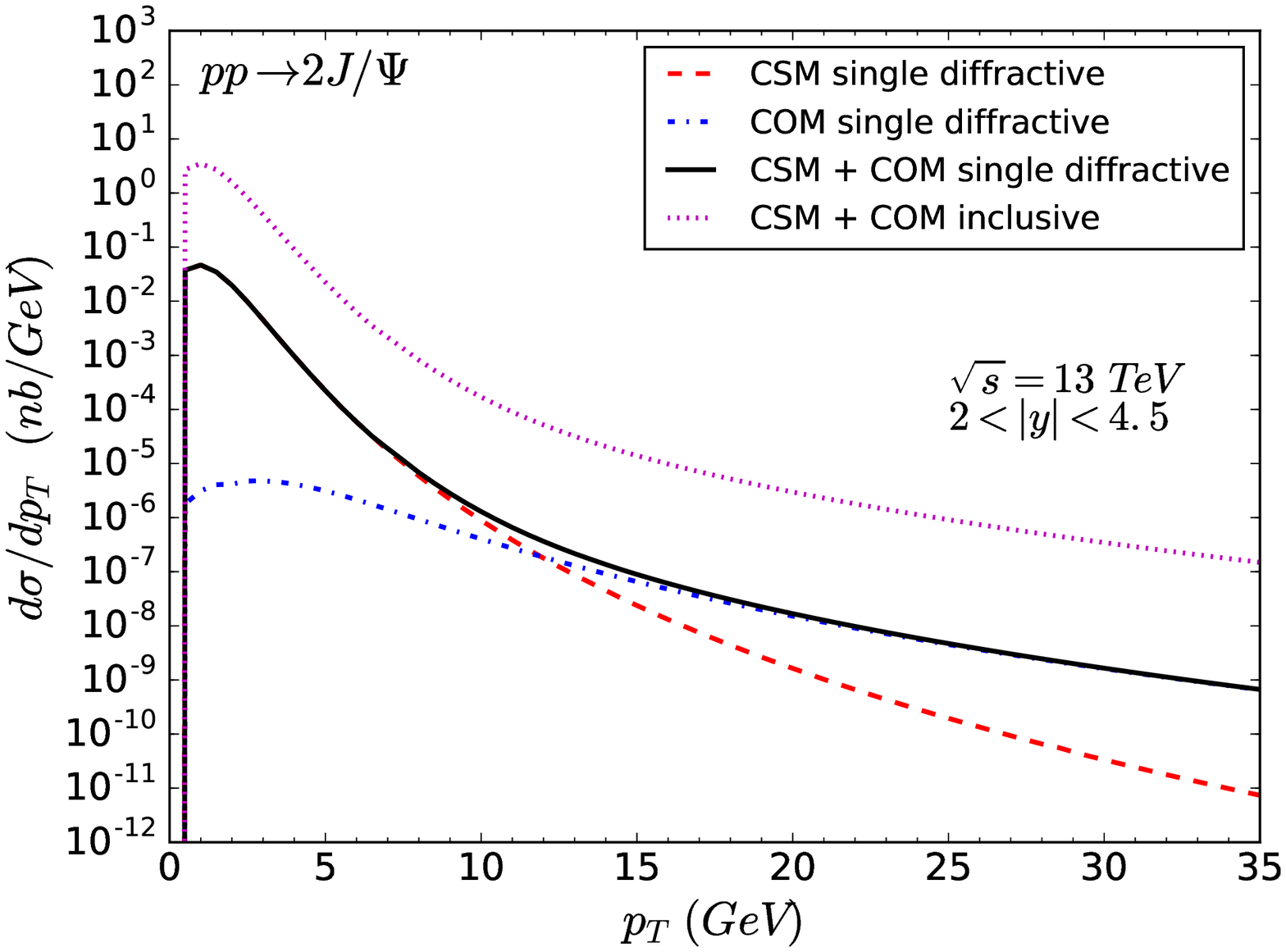}
}
{
\includegraphics[scale = .4]{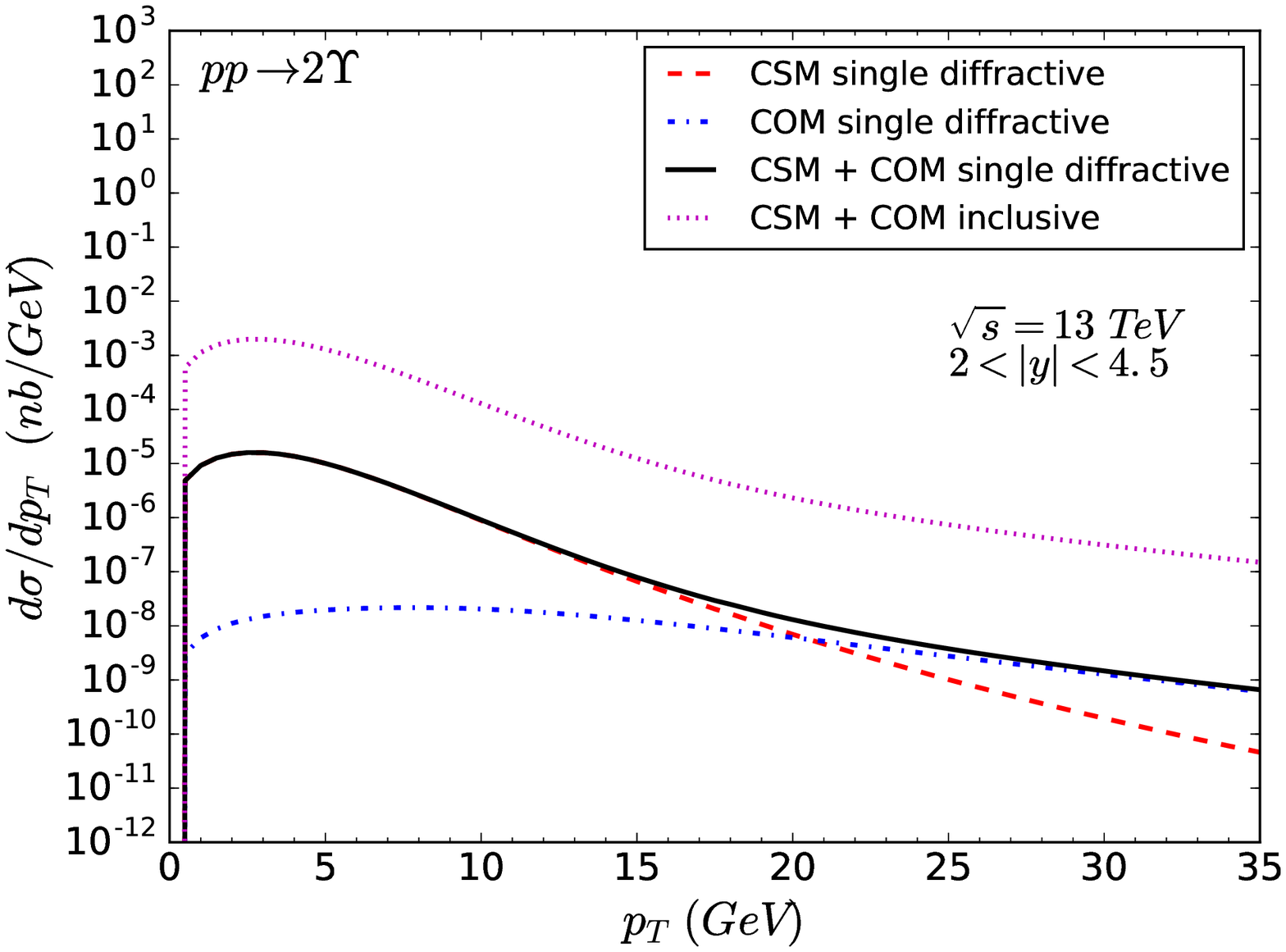}
}}
\end{tabular}
\caption{Transverse momentum distributions for the double $J/\Psi$ (left panels) and double $\Upsilon$ (right panels) production in single diffractive processes considering the full rapidity range (upper panels) and the LHCb range (lower panels).} 
\label{Fig:SD}
\end{figure}

\section{Results}
\label{res}
In what follows we will present our predictions for the single and double diffractive production of $J/\Psi J/\Psi$ and $\Upsilon \Upsilon$ in $pp$ collisions at the Run 2 LHC  energy ($\sqrt{s}=13$ TeV). We will estimate the rapidity ($y$) and transverse momentum ($p_T$) distributions for these processes and a comparison with the inclusive predictions will also be presented. In the case of double diffractive processes, these distributions can be estimated directly from the differential cross section given by
\begin{eqnarray}
\frac{d\sigma }{dydp_T^2}=  \langle |S|^2\rangle \cdot \int_{x_{1\, min}} dx_1
{g^D(x_1,\mu^2)g^D(x_2,\mu^2)}\frac{x_1x_2}{2x_1-\bar{x}_Te^y}
\sum_{n=1,8} \frac{d\hat{\sigma}}{d\hat{t}}[gg\rightarrow
 Q\bar{Q}_n(^3S_1) Q\bar{Q}_n(^3S_1)]
\cdot \langle {\cal{O}}_n^{H}(^3S_1) \rangle^2  \,\,,
\label{csdif2}
\end{eqnarray}
where 
$x_{1\, min}=\frac{\bar{x}_Te^{y}}{2-\bar{x}_Te^{-y} }$, 
$x_2=\frac{x_1\bar{x}_Te^{-y}}{2x_1-\bar{x}_Te^y}$, 
$\bar{x}_T=\frac{2m_T}{\sqrt{s}}$ and $m_T=\sqrt{M^2+p_T^2}$. The Eq. (\ref{csdif2}) can be directly generalized for single diffractive and inclusive processes by the adequate replacing of the diffractive gluon distribution by the inclusive one.

\begin{figure}[t]
\begin{tabular}{c c}
\centerline{
{
\includegraphics[scale = .4]{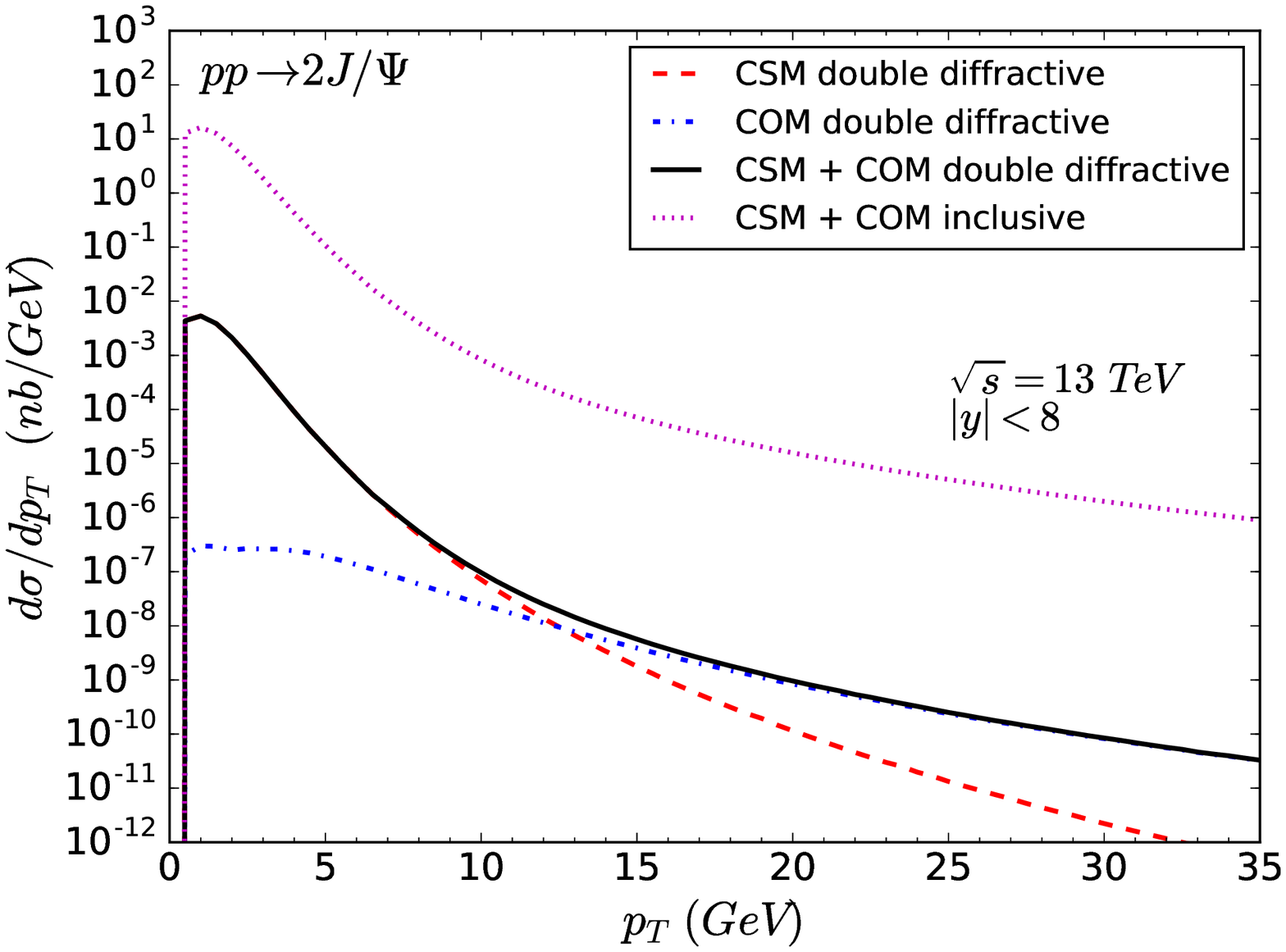}
}
{
\includegraphics[scale = .4]{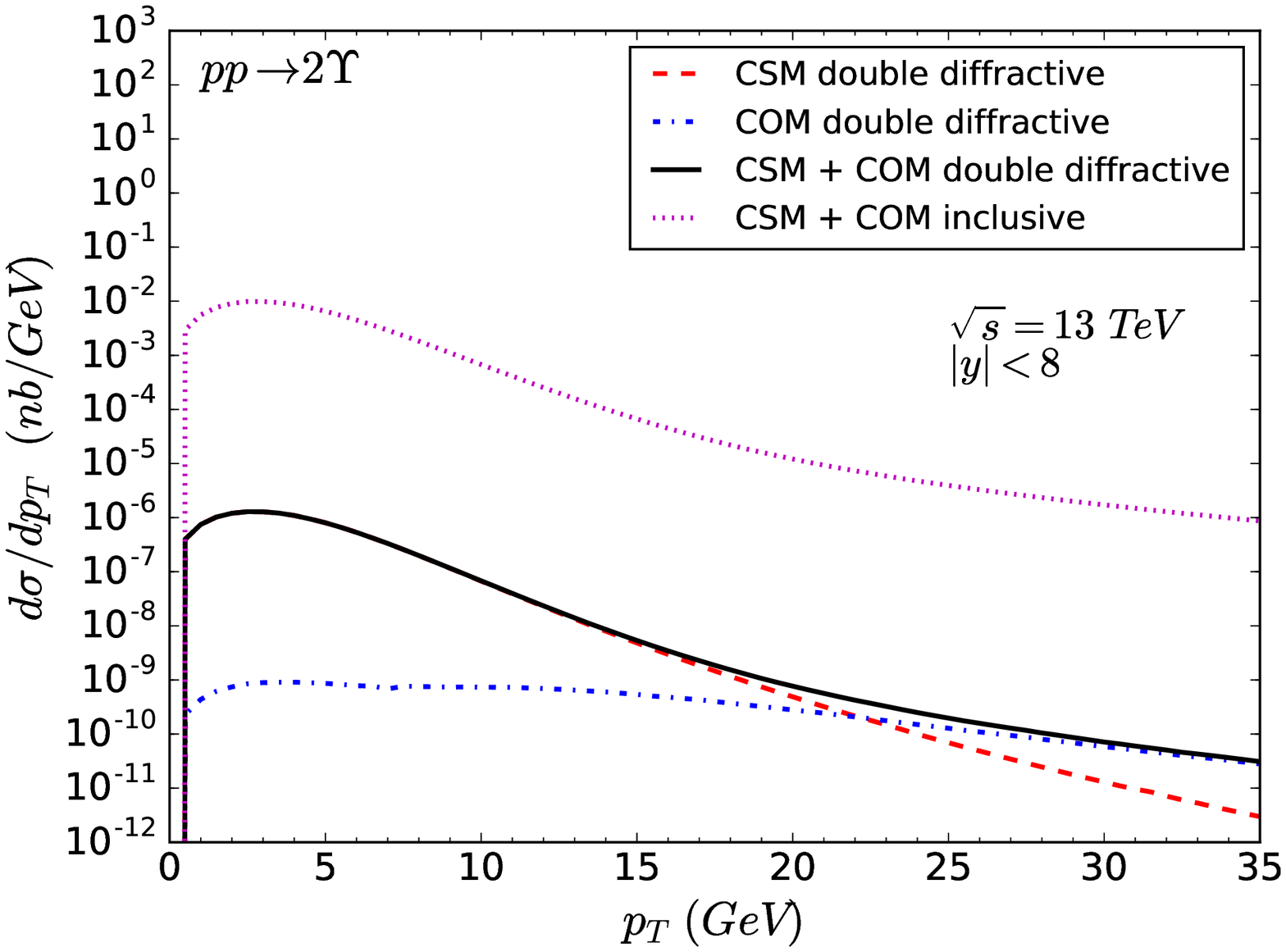}
}}\\
\centerline{
{
\includegraphics[scale = .4]{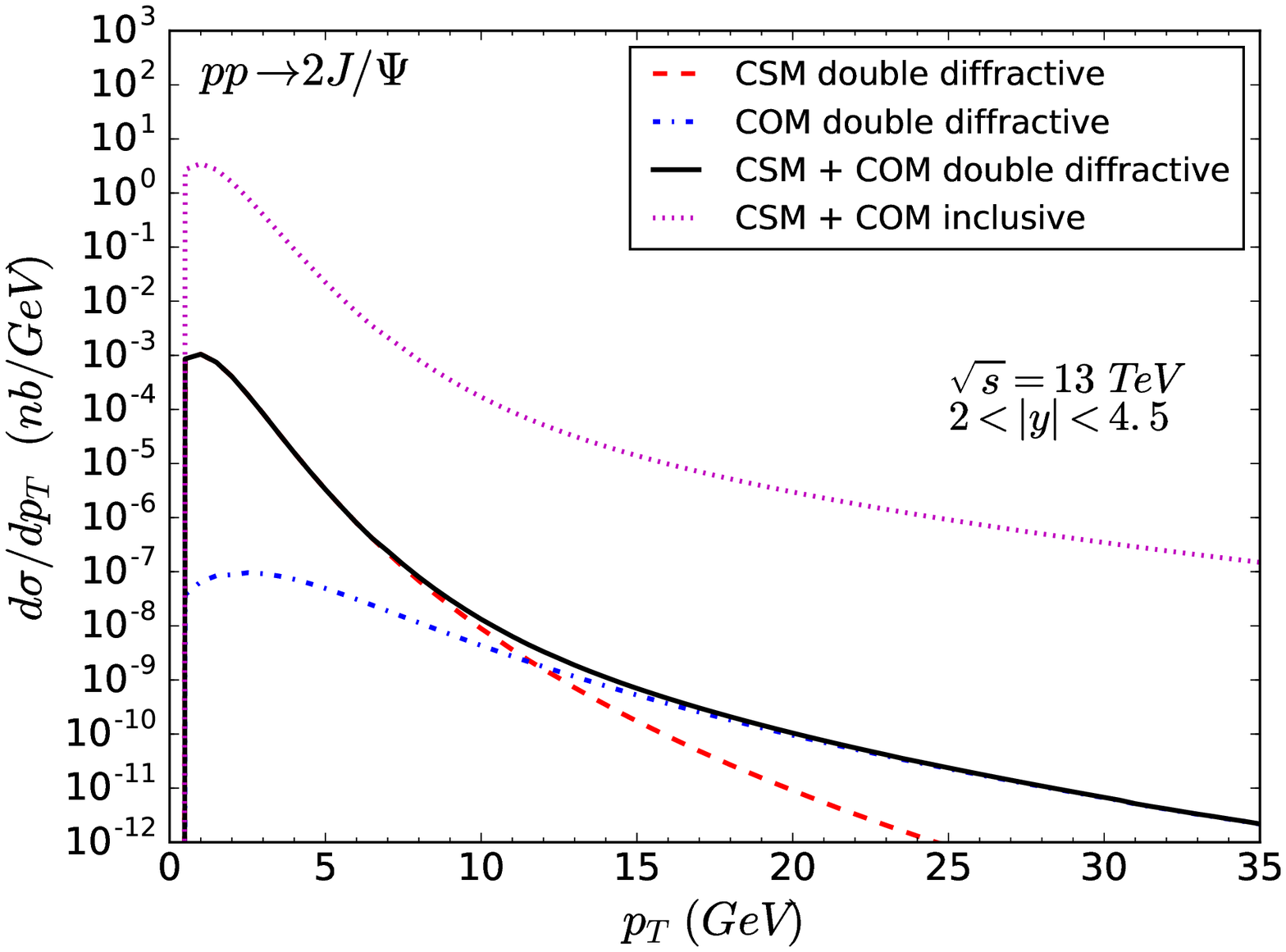}
}
{
\includegraphics[scale = .4]{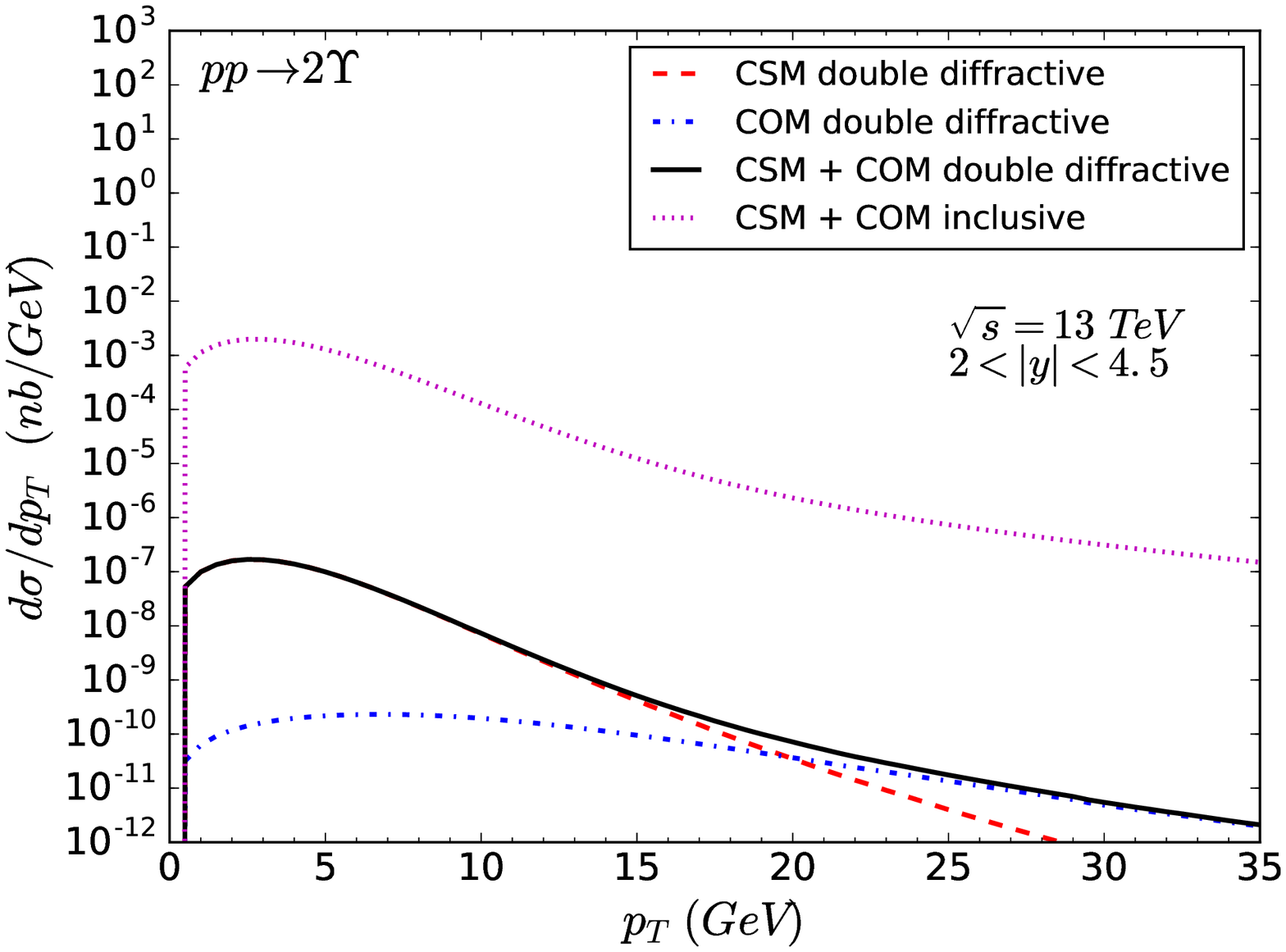}
}}
\end{tabular}
\caption{Transverse momentum distributions for the double $J/\Psi$ (left panels) and double $\Upsilon$ (right panels) production in double diffractive processes considering the full rapidity range (upper panels) and the LHCb range (lower panels).} 
\label{Fig:DD}
\end{figure}

In Fig. \ref{Fig:Rapidez} we present our predictions for the rapidity distributions for the  double $J/\Psi$ (left panel) and double $\Upsilon$ (right panel) production in single and double diffractive interactions. The inclusive { predictions} are also presented for comparison. In our calculations we have included the color-{ singlet} and color-octet mechanisms, which are denoted, respectively, by CSM and COM hereafter. 
The rapidity distributions are flat at the central rapidity region ($y \approx 0$), with the inclusive prediction being  a factor $ \approx 30$ ($10^3$) larger than SD (DD) one for the double $J/\Psi$ production. On the other hand, for the double $\Upsilon$ production, we predict that the inclusive prediction is a factor $ \approx 10^2$ ($10^4$) larger than SD (DD) result. These differences are also observed in the predictions for the total cross sections, presented in Table \ref{tabela} for two rapidity ranges. In particular, we present our predictions for the kinematical range probed by the LHCb Collaboration. In this case, the single and double diffractive predictions are reduced by approximately one order of magnitude in comparison to the full rapidity range. 
Our predictions for the double $J/\Psi$ production in the LHCb range for DD interactions are similar to those obtained in Ref. \cite{khoze} for the exclusive production.  As already emphasized in the Introduction, the topology of the final state of these two processes is different. While in central exclusive processes one has only the leading hadrons, two quarkonia and nothing else, in the double diffractive production one expect to have some extra particles coming from the Pomeron remnants. Consequently, in principle, the experimental separation between these two processes can be performed at smaller luminosities, as those presented in the LHCb detector. If it is feasible, a future analysis of the diffractive events can be useful to constrain the underlying assumptions present in description of these processes.  In particular, it will allow to improve the treatment of the absorptive effects that lead to the  breakdown of factorization in diffractive processes at hadronic colliders. On the other hand, if the separation of the exclusive and diffractive events is not possible due to large pile-up, our results indicate that the background associated to diffractive interactions cannot be disregarded in the selection of the exclusive events.

\begin{figure}[t]
\begin{tabular}{c c}
\centerline{
{
\includegraphics[scale = .4]{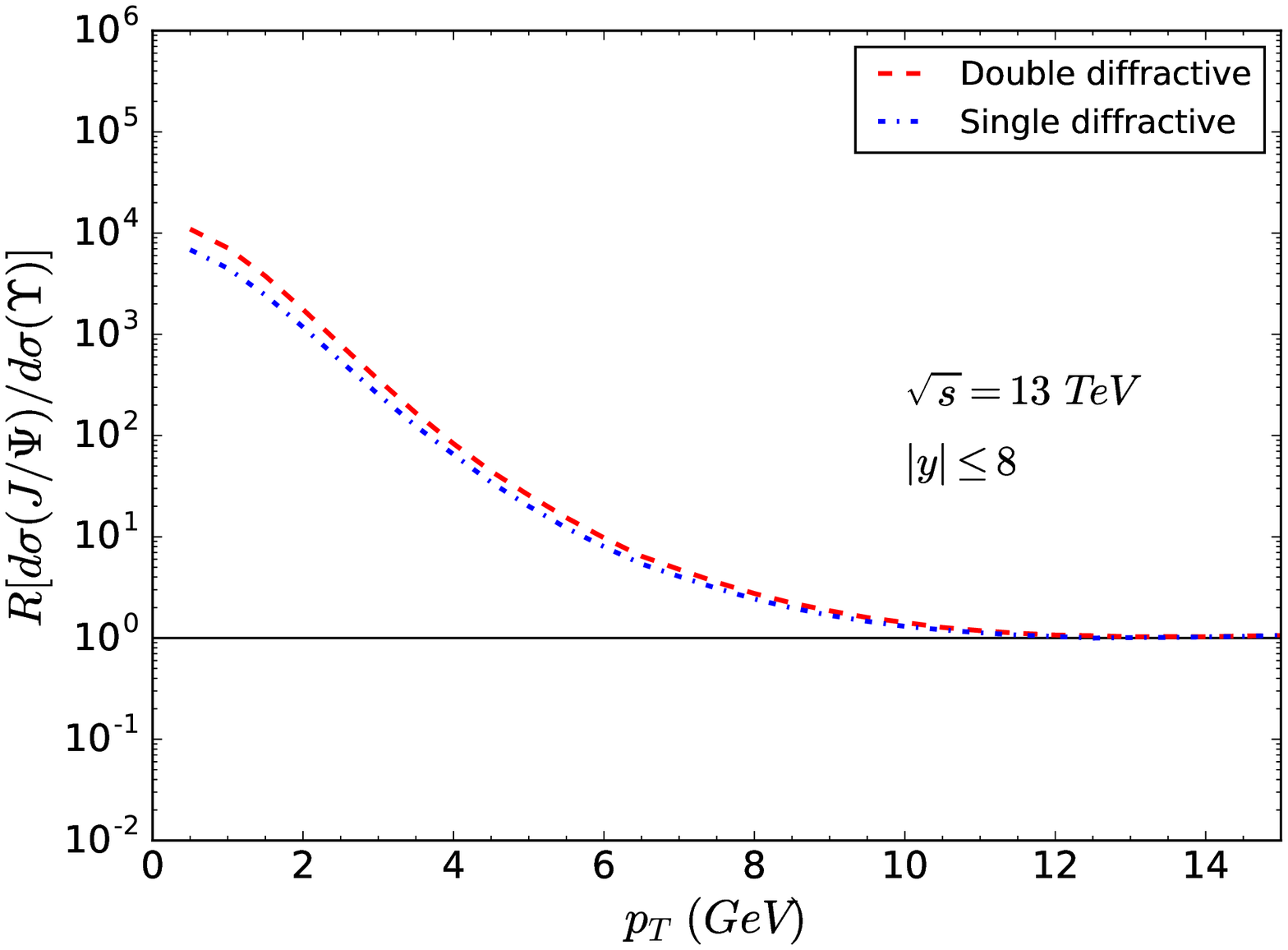}
}
{
\includegraphics[scale = .4]{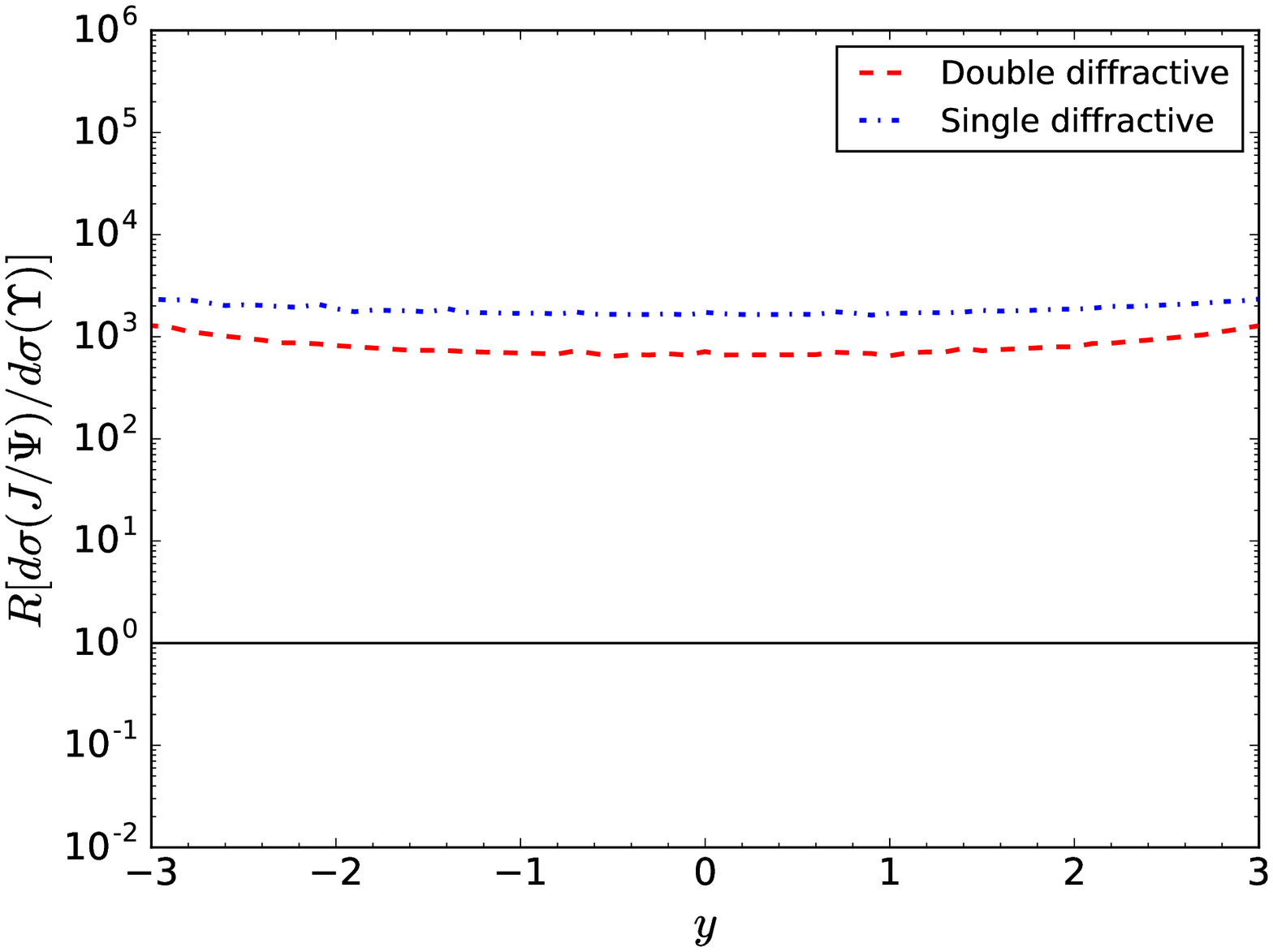}
}} 
\end{tabular}
\caption{Transverse momentum (left panel) and rapidity (right panel) dependencies of the ratio between the double $J/\Psi$ and double $\Upsilon$ cross sections for single and double diffractive processes.} 
\label{Fig:ratio}
\end{figure}

In Fig. \ref{Fig:SD} we present our predictions for the transverse momentum distributions for the double $J/\Psi$ (left panels) and double $\Upsilon$ (right panels) production in single diffractive processes considering the full rapidity range (upper panels) and the LHCb range (lower panels). The { contributions} of the color-singlet and color-octet mechanisms are presented separately, as well as the sum of both for the single diffractive and inclusive production. The shape of the $p_T$ - distributions for single diffractive and inclusive processes are very similar, with the distribution vanishing for $p_T \rightarrow 0$, in agreement with the results obtained in Ref. \cite{Ko}. For the full rapidity range (upper panels), one have that the contribution of the color-singlet mechanism is dominant at small values of $p_T$. It implies that the magnitude of the total cross sections { is} determined by this contribution and are not affected by the current uncertainty in the determination of the color-octet matrix elements. On the other hand, our results indicate that the color-octet mechanism determines the behaviour of the distribution  for $p_T \ge 10$ (20) GeV for the double $J/\Psi$  ($\Upsilon$) production. We obtain similar results for the LHCb range (lower panels), with the main difference being the normalization.

The predictions for double diffractive processes are presented in Fig. \ref{Fig:DD}. As in the SD case, the behaviour of the distribution at small - $p_T$ is determined by the color-singlet  contribution and the color-octet one only contributes at large transverse momentum. Such result is expected, since the single and double diffractive processes are determined by the same differential cross section for the subprocesses.

As discussed before, in our calculations we are assuming that the contribution of the absorptive effects can be factorized and { taken} into account by the multiplicative factor 
$\langle |S|^2\rangle$, which is assumed to be a constant { that is} independent { on} the final state produced in the diffractive interaction and the kinematical range considered. Such strong assumption can be tested by a future analysis of the ratio between the cross sections for the double $J/\Psi$ and double $\Upsilon$ production in single and double diffractive processes. If the assumption is correct, the transverse momentum and rapidity behaviours of the ratio will be independent of $\langle |S|^2\rangle$. Additionally, the impact of the next-to-leading order corrections and the dependence on the modelling of the inclusive and diffractive gluon distributions are expected to cancel in the ratio.  Moreover, considering that the cross sections are dominated by the color-singlet contribution, which is reasonably well known, and that the inclusive and diffractive gluon distributions are also well determined in the kinematical range of interest ($x \approx 10^{-3}$), a future analysis of the ratio can also be considered a direct probe of the framework considered to describe the double quarkonium production.  
In Fig. \ref{Fig:ratio} we present our predictions for the transverse momentum (left panel) and rapidity (right panel) dependencies of the ratio. The predictions for the single and double diffractive production are similar. At large - $p_T$ ($\ge 10$ GeV) we predict that the ratio will be $\approx 1$. Moreover, our results indicate that the ratio is almost rapidity independent. Similar results are obtained considering the LHCb kinematical range. If a different behaviour is observed in  a future experimental analysis, it will give us some hint about the treatment and kinematical dependence of the  soft gluon interactions in diffractive interactions. 

{
Some comments are in order before to summarize our main results in the next Section. In our calculations we have considered that the renormalization and factorization scales are equal to the transverse mass. As demonstrated in previous studies of the inclusive double quarkonium production \cite{lans4,lans2,lans3}, the predictions  are strongly sensitive to this choice, since the cross section at leading order is proportional to $\alpha_s^4$. As the dependence on the partonic cross sections is the same for inclusive and diffractive processes, we expect that a similar behaviour is also present in our predictions. In particular, a variation of the renormalization and factorization scales up and down by a factor 2 with respect to the central value used in our calculations is expected to modify our predictions by  $\approx 70 \%$ \cite{lans4}. Such uncertainty can be reduced in the future by precise measurements of the inclusive production. Another input parameter entering our calculations is the nonperturbative wave function of the quarkonium at the origin $R_H(0)$. As our predictions are dominated by the color singlet contributions, we have that if a different value for $R_H(0)$ is chosen, the new predictions can be obtained from the results presented here by multiplying our predictions by a factor of $[|R_H(0)|^2_{new}/|R_H(0)|^2]^2$, where $|R_H(0)|^2_{new}$ is the new value for the square of the radial wave function at the origin. Finally, in our analysis we have considered the CTEQ6L parametrization for the inclusive gluon distribution and the fit B of the H1 parametrization for the diffractive gluon distribution. As the main contribution for the cross sections come from values of $x \approx 10^{-3}$, where the inclusive gluon distribution is well determined, with different pametrizations predicting similar values, the impact of use a different model is small. In the diffractive case, we have verified that our predictions are modified by $\lesssim 9 \%$ if the fit A is used as input in the calculations. All these aspects imply an uncertainty of order of a factor $\lesssim 2$ in our predictions, which can be strongly reduced when the ratio of cross sections is considered. 
}

\section{Summary}
\label{conc}

The experimental analysis of the double quarkonium production became a reality during the last years. These results have attracted a renewed attention to the theoretical description of this process, which is expected to 
provide important insights that will allow us to improve our understanding of the hadron structure and the production mechanism. Moreover, the recent installation of forward detectors can be considered the begin  of a new era in the study of Diffractive Physics,  whose description is still one of the great challenges of the QCD. Consequently, the study of the double heavy quarkonium in diffractive processes 
has the potentiality of provide results that can be used to improve the understanding of several aspects in the theory of  strong interactions. That was the main motivation for the study performed in this paper, where we have presented a comprehensive analysis of the double heavy quarkonium { production} in single and double diffractive processes. We have provided predictions for the transverse momentum and rapidity distributions and for the total cross sections, derived using the NRQCD formalism for the quarkonium production and the Pomeron Resolved Model for the description of the diffractive interaction. We have demonstrated that the cross sections are dominated by the color-{ singlet} mechanism, with the color-octet one being important only at large - $p_T$, where the { magnitude} of the cross sections is strongly reduced. Moreover, our results indicate that the contribution of the diffractive production is non-negligible at the Run 2 LHC energy, which implies that a future experimental analysis is, in principle, feasible. We also have indicated that the analysis of { the} ratio between cross sections can be useful to probe the treatment of the absorptive corrections. Finally, our results indicate that a future analysis of the diffractive events can be useful to constrain the underlying assumptions present in { the} description of the double quarkonium production.

\begin{acknowledgments}
VPG acknowledge useful discussions with  M. Rangel and R. McNulty. This work was  partially financed by the Brazilian funding agencies CNPq, CAPES,  FAPERGS and  INCT-FNA (process number 464898/2014-5).
\end{acknowledgments}

\hspace{1.0cm}

\end{document}